\documentclass[12pt]{article}

\usepackage{fancybox}

\usepackage{cite}
\usepackage{float}
\usepackage{amsfonts}
\usepackage{amsmath}
\usepackage{amsbsy}
\usepackage{graphicx}
\usepackage{amssymb}
\usepackage{amsthm}
\usepackage{bm}
\usepackage{epsfig}
\usepackage{latexsym}
\usepackage{pdflscape}
\usepackage{color}
\usepackage{here}
\usepackage{graphicx}
\numberwithin{equation}{section}

\allowdisplaybreaks

\DeclareMathOperator{\tr}{tr}
\DeclareMathOperator{\Res}{Res}

\newcounter{aff}

\begin{document}

\begin{titlepage}
\begin{flushright}
{\footnotesize OCU-PHYS 432}
\end{flushright}
\bigskip
\begin{center}
{\LARGE\bf Instanton Effects in Orientifold ABJM Theory}\\
\bigskip\bigskip
{\large 
Sanefumi Moriyama\footnote{\tt moriyama@sci.osaka-cu.ac.jp}
\quad and \quad
Takao Suyama\footnote{\tt suyama@sci.osaka-cu.ac.jp}
}\\
\bigskip
${}^*${\it Department of Physics, Graduate School of Science,
Osaka City University}\\
${}^\dagger${\it Osaka City University 
Advanced Mathematical Institute (OCAMI)}\\
{\it 3-3-138 Sugimoto, Sumiyoshi, Osaka 558-8585, Japan}
\end{center}

\bigskip

\begin{abstract}

We investigate another supersymmetric Chern-Simons theory called the orientifold ABJM theory, which replaces the unitary supergroup structure of the ABJM theory with an orthosymplectic one.
Its non-perturbative structure is completely clarified by considering the duplication of the quiver.

\bigskip

\centering\includegraphics[scale=0.75,angle=-90]{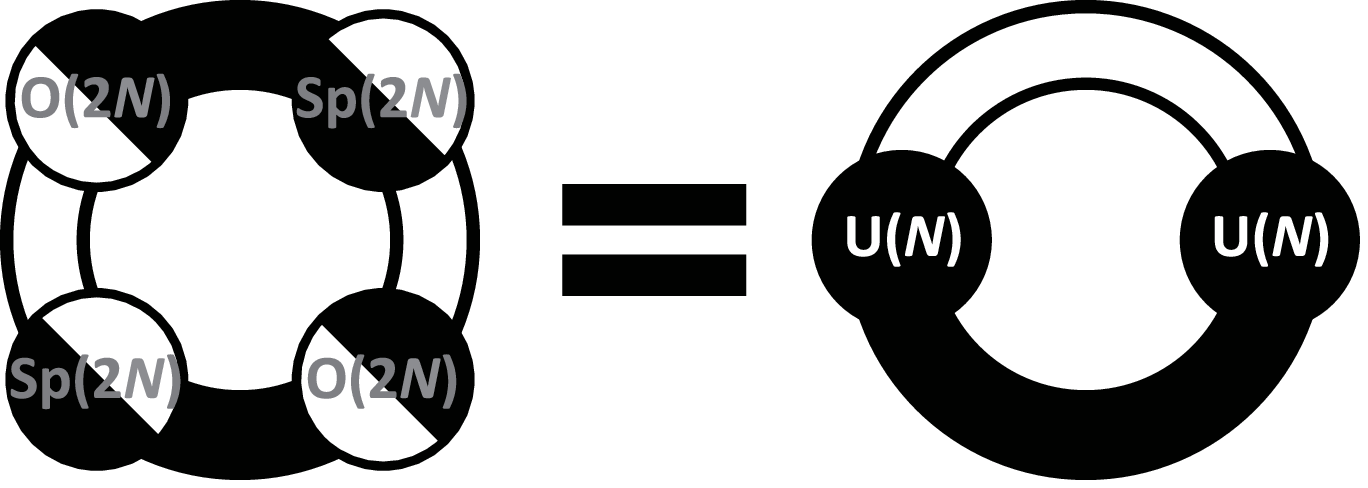}

\end{abstract}
\end{titlepage}

\tableofcontents

\section{Introduction}\label{introduction}
Recently, there has been much progress in understanding the partition function of supersymmetric Chern-Simons theories.
The first of such theories was, of course, the ${\cal N}=6$ supersymmetric one called ABJM theory \cite{ABJM}, which has the gauge group U$(N)_k\times$U$(N)_{-k}$ (with the subscripts denoting the Chern-Simons levels) and two pairs of bifundamental matters between the two U$(N)$ factors.
This theory is the worldvolume theory of multiple M2-branes placed on a background ${\mathbb C}^4/{\mathbb Z}_k$.
Using the localization theorem, the partition function $Z_k(N)$ on $S^3$ is reduced to a matrix model \cite{KWY}, where the resulting expression can be interpreted in terms of the hidden supergroup U$(N|N)$ \cite{DT}.
After the perturbative terms of the partition function $Z_k(N)$ were summed to obtain an Airy function \cite{FHM} following the study of the large $N$ behavior $N^{\frac{3}{2}}$ in \cite{DMP1,DMP2}, the Fermi gas formalism was proposed \cite{MP}, where the Airy function is easily rederived in the disguise of a cubic polynomial in the grand canonical ensemble.
Switching to the grand potential $J_k(\mu)$,
\begin{align}
\sum_{n=-\infty}^\infty e^{J_k(\mu+2\pi in)}
=\sum_{N=0}^\infty|Z_k(N)|e^{\mu N},
\label{totalgrandpot}
\end{align}
defined with the oscillatory terms, which originates from the trivial $2\pi i$ shift of $\mu$, properly subtracted \cite{HMO2}, we found that the grand potential has both perturbative and non-perturbative contributions,
\begin{align}
J_k(\mu)=J^\text{(pert)}_k(\mu_\text{eff})
+J^\text{(np)}_k(\mu_\text{eff}),
\end{align}
where the result is simplified using the redefined effective chemical potential $\mu_\text{eff}=\mu+{\cal O}(e^{-2\mu})$.
In addition to the perturbative part,
\begin{align}
J^\text{(pert)}_k(\mu_\text{eff})
=\frac{C_k}{3}\mu_\text{eff}^3+B_k\mu_\text{eff}+A_k,
\end{align}
giving rise to the Airy function, we were able to study the non-perturbative effects explicitly \cite{HMO2,CM,HMO3,HMMO}, which turned out to be expressed in terms of the free energy of topological strings on local ${\mathbb P}^1\times{\mathbb P}^1$ ($s_L=2j_L+1$, $s_R=2j_R+1$),
\begin{align}
&J^\text{(np)}_k(\mu_\text{eff})
=\sum_{j_L,j_R}\sum_{{\bf d}=(d^1,d^2)}
N^{\bf d}_{j_L,j_R}\nonumber\\
&\quad\times
\sum_{n=1}^\infty
\biggl[\frac{s_R\sin 2\pi g_sns_L}{n(2\sin\pi g_sn)^2\sin 2\pi g_sn}
e^{-n{\bf d}\cdot{\bf T_\text{eff}}}
+\frac{\partial}{\partial g_s}
\biggl(g_s\frac{-\sin\frac{\pi n}{g_s}s_L\sin\frac{\pi n}{g_s}s_R}
{4\pi n^2(\sin\frac{\pi n}{g_s})^3}
e^{-\frac{n{\bf d}\cdot{\bf T_\text{eff}}}{g_s}}
\biggr)\biggr].
\label{even}
\end{align}
Here the K\"ahler parameters $T^1_\text{eff},T^2_\text{eff}$ and the coupling constant $g_s$ are identified with the chemical potential and the inverse level by
\begin{align}
T^1_\text{eff}=\frac{4\mu_\text{eff}}{k}+\pi i,\quad
T^2_\text{eff}=\frac{4\mu_\text{eff}}{k}-\pi i,\quad
g_s=\frac{2}{k},
\label{ABJMTgs}
\end{align}
with the $B$-field effect $\pm\pi i$, and the overall integral coefficient $N^{\bf d}_{j_L,j_R}$ is the BPS index.
It can be proved that $N^{\bf d}_{j_L,j_R}$ is non-vanishing only when $2j_L+2j_R-1$ is even (see e.g.~\cite{HMMO}).
It is then interesting to ask how the grand potential is constrained so that it reduces to the free energy of topological strings.
We are also interested in whether there are any other theories that have the same topological string expression and whether there are any expressions other than the topological one that can appear as the grand potential in the supersymmetric Chern-Simons theories.

To answer these questions, some generalizations may be helpful.
The ability to make generalizations depends on the number of supersymmetries.
Basically, the more supersymmetries the theory possesses, the more solvable the theory is.
The ${\cal N}=3$ supersymmetric theories were constructed for all gauge groups and all matters.
Among them, the $\widehat A_M$ quiver gauge theories with the gauge group $\prod_{a=1}^{M+1}$U$(N)_{k_a}$ (having the vanishing total level $\sum_ak_a=0$) and pairs of the bifundamental matters cyclically connecting two adjacent factors U$(N)_a$ and U$(N)_{a+1}$ were proposed.
It was found that when the levels are expressed as $k_a=(k/2)(s_a-s_{a-1})$ with $s_a=\pm 1$, the theories have enhanced ${\cal N}=4$ supersymmetries \cite{IK4}.

Two major examples of the generalizations are the orbifold ABJM theory and the $(q,p)$ model.
The $(q,p)$ model \cite{MN1} is described by a quiver with $q$ sequential edges of $s_a=+1$ and $p$ edges of $s_a=-1$.
In particular, it was exciting to find that the grand potential of the $(2,2)$ model takes the same topological string expression \eqref{even} with the diagonal Gopakumar-Vafa invariant being that of the $D_5$ del Pezzo geometry \cite{MN3}.
See \cite{HHO} for further intensive studies on general values of $(q,p)$.

In terms of the quiver diagram, the orbifold ABJM theory is an $r$-times repetition of the original ABJM theory \cite{GW,HLLLP1}.
In \cite{HM} (see also \cite{OWZ}) it was found that as long as the new quiver is an $r$-times repetition of the original quiver, the grand potential of the new theory $J^{[r]}_k(\mu)$ is expressed in terms of that of the original theory $J^{[1]}_k(\mu)$ by
\begin{align}
e^{J^{[r]}_k(\mu)}
=\sum_{\sum_jn_j=0}\prod_{j=-\frac{r-1}{2}}^{\frac{r-1}{2}}
e^{J^{[1]}_k(\frac{\mu+2\pi ij}{r}+2\pi in_j)}.
\label{repetitive}
\end{align}
More explicitly, for $r=2$, we have
\begin{align}
&J^{[2]}_k(\mu)=J^{[1]}_k\Bigl(\frac{\mu-\pi i}{2}\Bigr)
+J^{[1]}_k\Bigl(\frac{\mu+\pi i}{2}\Bigr)\nonumber\\
&\quad
+\log\biggl[1+\sum_{n\ne 0}
e^{J^{[1]}_k(\frac{\mu-\pi i}{2}+2\pi in)
+J^{[1]}_k(\frac{\mu+\pi i}{2}-2\pi in)
-J^{[1]}_k(\frac{\mu-\pi i}{2})
-J^{[1]}_k(\frac{\mu+\pi i}{2})}\biggr].
\label{2repeat}
\end{align}
Here, we usually project out some terms in the first line but we need to add the second line to preserve the $2\pi i$ shift symmetry.
In this sense, we will often refer to the first line with $n=0$ as the untwisted sector and the second line with $n\ne 0$ as the twisted sectors using the terminology of the string orbifold theory.
This analogy works even for $r>2$.
Although the expression \eqref{repetitive} appears to contain infinite terms, the summation terminates at a finite $n$ for a fixed instanton number because the exponent of each twisted $(n\ne 0)$ sector always contains a decaying factor originating from the perturbative terms.

In this paper, we will further explore generalizations of the ABJM theory.
Although the progress is guided by the number of supersymmetries, as we have already mentioned, there remains a theory with a large number of supersymmetries for which the discussion of its non-perturbative effects is still missing: the ${\cal N}=5$ orientifold ABJM theory \cite{HLLLP2,ABJ}.
This theory can be realized in a type IIB setup as the worldvolume theory of a brane system by adding an O3$^-$-plane in parallel with the original D3-branes used to realize the ABJM theory.
In the dual M-theory description, this theory describes multiple M2-branes in M-theory on a D-type orbifold \cite{HLLLP2,ABJ}.
Again, after the supersymmetric localization, the partition function is reduced to a matrix model.
Instead of the previous interpretation in terms of the supergroup U$(N|N)$ for the ABJM theory, this theory can be interpreted in terms of the supergroup OSp$(2N|2N)$.
Hereafter, we will call this theory the orthosymplectic (OSp) theory.

\begin{figure}
\centering\includegraphics[scale=0.75,angle=-90]{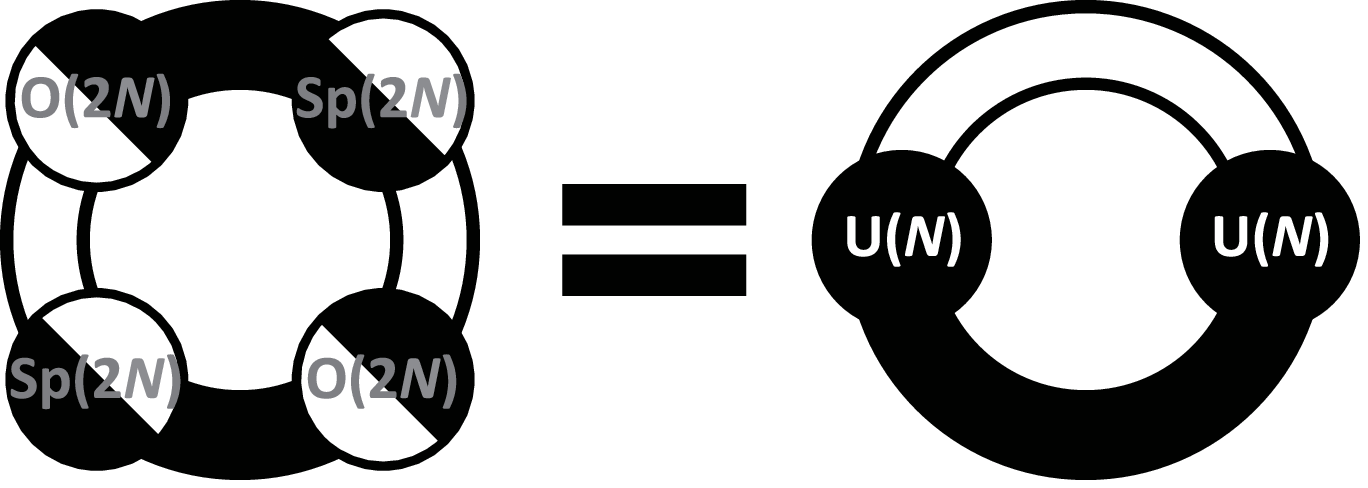}
\caption{Schematic relation between the grand potential of the OSp theory and that of the ABJM theory.}
\label{fromosptoabjm}
\end{figure}

We follow the recent progress in establishing the Fermi gas formalism and compute exact values of the partition function $Z_k(N)$ for various values of $k$ and $N$.
Surprisingly, when we compare the resultant grand potential with that of the ABJM theory, we encounter the same structure as \eqref{repetitive}.
Namely, we find that the grand potential of the OSp theory is related to that of the ABJM theory by
\begin{align}
J^\text{OSp[2]}_{k}(\mu)=J^\text{ABJM[1]}_k(\mu/2)-\log 2.
\label{OSpABJM}
\end{align}
Here $J^\text{OSp[2]}_k(\mu)$ is the grand potential of a duplicate OSp theory in the sense of \eqref{2repeat}, while $J^\text{ABJM[1]}_k(\mu)=J^\text{ABJM}_k(\mu)$ is the grand potential of the original ABJM theory.
Since we already know $J^\text{ABJM}_k(\mu)$, this relation almost fixes the grand potential of the OSp theory.
In this sense, we claim that the partition function of the OSp theory on $S^3$ is solved.
See figure \ref{fromosptoabjm} for a schematic expression and section \ref{fromto} for further explanation.

The organization of this paper is as follows.
In the next section, we will study the OSp matrix model in full detail, first presenting the Fermi gas formalism and then analyzing it.
Then, we conclude with some discussion on future directions.
The appendix is devoted to the integration formulas necessary for the analysis of the OSp matrix model.

\section{OSp matrix model}
Let us start our analysis of the partition function of the OSp theory.
This theory was proposed in \cite{HLLLP2,ABJ} and possesses the enhanced ${\cal N}=5$ supersymmetries.
The gauge group of the theory is O$(2N)\times$USp$(2N)$, whose Chern-Simons levels are $k$ and $k/2$, respectively, in the convention of \cite{GHN}\footnote{
Note that the parameter $k$ appearing in \cite{HLLLP2} is different from $k$ in this paper, $k^\text{here}=2k^\text{HLLLP}$, owing to the difference in the normalization of the trace. 
}.
According to \cite{GHN}, $k$ must be an even integer.

After applying the localization technique \cite{KWY}, the partition function on $S^3$ is given by\footnote{We assume that the level $k$ is positive.
For negative $k$, we can simply consider its complex conjugate.}
\begin{align}
Z_k(N)=\int\frac{D^N\mu}{N!}\frac{D^N\nu}{N!}
\frac{V_\text{O}V_\text{Sp}}{H},
\label{Z_k(N)}
\end{align}
with the integration originating from the tree-level contribution
\begin{align}
D\mu_i=\frac{d\mu_i}{4\pi}e^{\frac{ik}{4\pi}\mu_i^2},\quad
D\nu_j=\frac{d\nu_j}{4\pi}e^{-\frac{ik}{4\pi}\nu_j^2},
\end{align}
and the measure originating from the one-loop contributions of the vector multiplets and the hypermultiplets
\begin{align}
V_\text{O}&=\prod_{i<j}^N\Bigl(2\sinh\frac{\mu_i-\mu_j}{2}\Bigr)^2
\Bigl(2\sinh\frac{\mu_i+\mu_j}{2}\Bigr)^2,\nonumber\\
V_\text{Sp}&=\prod_{i<j}^N\biggl(2\sinh\frac{\nu_i-\nu_j}{2}\Bigr)^2
\Bigl(2\sinh\frac{\nu_i+\nu_j}{2}\Bigr)^2
\prod_j^N\bigl(2\sinh\nu_j\bigr)^2,\nonumber\\
H&=\prod_{i,j}^{N}\Bigl(2\cosh\frac{\mu_i-\nu_j}{2}\Bigr)^2
\Bigl(2\cosh\frac{\mu_i+\nu_j}{2}\Bigr)^2.
\end{align}
Note that the integration measure is closely related to the invariant measure of OSp$(2N|2N)$ in the same way that the integration measure of the ABJM theory is related to U$(N|N)$.

\subsection{Fermi gas formalism}
Now let us propose a Fermi gas formalism for the OSp matrix model.
The Fermi gas formalism for this matrix model was first studied in \cite{MePu} and many important tools have already been given in \cite{HMO1}.

Using the Cauchy determinant formula 
\begin{align}
&\det\frac{1}{2(\cosh\mu_i+\cosh\nu_j)}
=\frac{\prod_{i<j}(2\sinh\frac{\mu_i-\mu_j}{2})
(2\sinh\frac{\mu_i+\mu_j}{2})
\prod_{i<j}(2\sinh\frac{\nu_i-\nu_j}{2})
(2\sinh\frac{\nu_i+\nu_j}{2})}
{\prod_{i,j}(2\cosh\frac{\mu_i-\nu_j}{2})
(2\cosh\frac{\mu_i+\nu_j}{2})},
\end{align}
we can rewrite the partition function (\ref{Z_k(N)}) as
\begin{align}
Z_k(N)&=\int\frac{D^N\mu}{N!}\frac{D^N\nu\prod_i(2\sinh\nu_i)^2}{N!}
\det\frac{1}{2(\cosh\mu_i+\cosh\nu_j)}
\det\frac{1}{2(\cosh\nu_j+\cosh\mu_i)}.
\end{align}
After expanding the determinant and rescaling the integration variables, we obtain
\begin{align}
Z_k(N)&=\frac{1}{N!}\sum_{\sigma\in S_N}
(-1)^\sigma\int\frac{d^N\mu}{(4\pi)^N}
\prod_{i=1}^N\int\frac{d\nu}{4\pi}
e^{\frac{i}{4\pi k}\mu_i^2}
\frac{1}{k}\frac{1}{\cosh\frac{\mu_i}{k}+\cosh\frac{\nu}{k}}
\nonumber \\
&\quad\times
e^{-\frac{i}{4\pi k}\nu^2}\Bigl(\sinh\frac{\nu}{k}\Bigr)^2
\frac{1}{k}\frac{1}{\cosh\frac{\nu}{k}+\cosh\frac{\mu_{\sigma(i)}}{k}}.
\label{Z_k(N)-2}
\end{align}

Recall that to rewrite the partition function of the ABJM matrix model into that of a Fermi gas system \cite{MP}, the phase space operators $\widehat q$ and $\widehat p$ satisfying $[\widehat q,\widehat p]=i\hbar$ and the coordinate eigenstate $|\mu\rangle$ normalized by $\langle\mu|\nu\rangle=2\pi\delta(\mu-\nu)$ were introduced.
Using them, we were able to express the entries of the determinant as
\begin{align}
\langle\mu|\frac{1}{2\cosh\frac{\widehat p}{2}}|\nu\rangle
=\frac{1}{k}\frac{1}{2\cosh\frac{\mu-\nu}{2k}},
\end{align}
with the identification $\hbar=2\pi k$.
To deal with the extra factor of the hyperbolic sine functions, the use of the following reflection operator and projectors was proposed in \cite{MePu}:
\begin{align}
\widehat R
=\int\frac{d\mu}{2\pi}|\mu\rangle\langle-\mu|,\quad
\widehat\Pi_\pm=\frac{1\pm\widehat R}{2}.
\end{align}
Then, using 
\begin{align}
\langle\mu|\frac{\widehat\Pi_+}{2\cosh\frac{\widehat p}{2}}|\nu\rangle
=\frac{1}{k}\frac{\cosh\frac{\mu}{2k}\cosh\frac{\nu}{2k}}
{\cosh\frac{\mu}{k}+\cosh\frac{\nu}{k}},\quad
\langle\mu|\frac{\widehat\Pi_-}{2\cosh\frac{\widehat p}{2}}|\nu\rangle
=\frac{1}{k}\frac{\sinh\frac{\mu}{2k}\sinh\frac{\nu}{2k}}
{\cosh\frac{\mu}{k}+\cosh\frac{\nu}{k}},
\label{Picosh}
\end{align}
we can rewrite the partition function (\ref{Z_k(N)-2}) as
\begin{align}
Z_k(N)&=\frac{1}{N!}\sum_{\sigma\in S_N}
(-1)^\sigma\int\frac{d^N\mu}{(2\pi)^N}\prod_{i=1}^N
\langle\mu_i|e^{-\widehat H}|\mu_{\sigma(i)}\rangle,
\label{Z_k(N)-3}
\end{align}
where the Hamiltonian is given by
\begin{align}
e^{-\widehat H}=e^{\frac{i}{2\hbar}\widehat q^2}
\biggl(\sinh\frac{\widehat q}{k}\biggr)^{-1}
\frac{\widehat\Pi_-}{2\cosh\frac{\widehat p}{2}}
\biggl(\sinh\frac{\widehat q}{k}\biggr)
e^{-\frac{i}{2\hbar}\widehat q^2}
\frac{\widehat\Pi_+}{2\cosh\frac{\widehat p}{2}}.
\label{twoproj}
\end{align}
Note that $\widehat R$ and $(2\cosh\frac{\widehat p}{2})^{-1}$ are commutative.
This is why the expressions in \eqref{Picosh} make sense.
The two projectors appearing in \eqref{twoproj} are redundant.
We can always combine them, but let us rewrite \eqref{twoproj} as
\begin{align}
e^{-\widehat H}=e^{\frac{i}{2\hbar}\widehat q^2}
\biggl(\sinh\frac{\widehat q}{k}\biggr)^{-1}
\frac{1}{2\cosh\frac{\widehat p}{2}}
\biggl(\sinh\frac{\widehat q}{k}\biggr)\widehat\Pi_+
e^{-\frac{i}{2\hbar}\widehat q^2}
\frac{\widehat\Pi_+}{2\cosh\frac{\widehat p}{2}}.
\label{hamiltonian}
\end{align}

The Hamiltonian looks considerably more difficult than that in the original ABJM theory.
However, a key observation for studying this Hamiltonian has already been proposed in \cite{HMO1}.
From the Baker-Campbell-Hausdorff formula, we can show
\begin{align}
\biggl(\sinh\frac{\widehat q}{k}\biggr)
\biggl(2\cosh\frac{\widehat p}{2}\biggr)
+\biggl(2\cosh\frac{\widehat p}{2}\biggr)
\biggl(\sinh\frac{\widehat q}{k}\biggr)=0.
\label{BCHresult}
\end{align}
Naively, it appears that using \eqref{BCHresult} we can cancel the unwanted hyperbolic sine functions to obtain $-1$ in the Hamiltonian \eqref{hamiltonian}.
It is not, however, as simple.
When we consider the cancellation, the following remaining term appears:
\begin{align}
\langle\mu|\Biggl[\biggl(\sinh\frac{\widehat q}{k}\biggr)^{-1}
\frac{1}{2\cosh\frac{\widehat p}{2}}
\biggl(\sinh\frac{\widehat q}{k}\biggr)
+\frac{1}{2\cosh\frac{\widehat p}{2}}\Biggr]|\nu\rangle
=\frac{1}{k}\frac{\sinh\frac{\mu+\nu}{2k}}{\sinh\frac{\mu}{k}}.
\label{noproj}
\end{align}
After combining \eqref{noproj} with the projector $\widehat\Pi_+$, the $\mu$ and $\nu$ dependences decouple,
\begin{align}
\langle\mu|\Biggl[\biggl(\sinh\frac{\widehat q}{k}\biggr)^{-1}
\frac{1}{2\cosh\frac{\widehat p}{2}}
\biggl(\sinh\frac{\widehat q}{k}\biggr)
+\frac{1}{2\cosh\frac{\widehat p}{2}}\Biggr]\widehat\Pi_+|\nu\rangle
=\frac{1}{k}\frac{1}{2\cosh\frac{\mu}{2k}}\cosh\frac{\nu}{2k}.
\end{align}
Since the right-hand side can be expressed as
\begin{align}
\frac{1}{k}\frac{1}{2\cosh\frac{\mu}{2k}}\cosh\frac{\nu}{2k}
=\langle\mu|\frac{1}{2\cosh\frac{\widehat p}{2}}|0\rangle
\langle\widetilde 0|\cosh\frac{\widehat q}{2k}|\nu\rangle,
\end{align}
with $\langle\widetilde 0|$ being the $\widehat p=0$ eigenstate
\begin{align}
\langle\widetilde 0|=\sqrt{k}\langle p=0|,
\end{align}
which is normalized such that
\begin{align}
\langle\widetilde 0|0\rangle=1,\quad
\langle\widetilde 0|\cosh\frac{\widehat q}{2k}|0\rangle=1,
\end{align}
after the cancellation we end up with two projectors,
\begin{align}
\biggl(\sinh\frac{\widehat q}{k}\biggr)^{-1}
\frac{1}{2\cosh\frac{\widehat p}{2}}
\biggl(\sinh\frac{\widehat q}{k}\biggr)\widehat\Pi_+
=-\frac{\widehat\Pi_+}{2\cosh\frac{\widehat p}{2}}
\biggl[1-|0\rangle\langle\widetilde 0|\cosh\frac{\widehat q}{2k}\biggr].
\label{scs}
\end{align}

Using \eqref{scs}, we can rewrite the Hamiltonian as
\begin{align}
e^{-\widehat H}=-e^{\frac{i}{2\hbar}\widehat q^2}
\frac{\widehat\Pi_+}{2\cosh\frac{\widehat p}{2}}
\biggl[1-|0\rangle\langle\widetilde 0|\cosh\frac{\widehat q}{2k}\biggr]
e^{-\frac{i}{2\hbar}\widehat q^2}
\frac{\widehat\Pi_+}{2\cosh\frac{\widehat p}{2}}.
\end{align}
Hence the grand canonical partition function
\begin{align}
\Xi_k(z)=\sum_{N=0}^\infty z^{N}Z_k(N)=\det(1+ze^{-\widehat H}),
\label{GCPF}
\end{align}
can be expressed as
\begin{align}
\Xi_k(z)&=\det\Biggl(1-ze^{-\frac{i}{2\hbar}\widehat q^2}
\frac{\widehat\Pi_+}{2\cosh\frac{\widehat p}{2}}
e^{\frac{i}{2\hbar}\widehat q^2}
\frac{\widehat\Pi_+}{2\cosh\frac{\widehat p}{2}}
\Biggr)
\nonumber\\&\qquad\times
\langle\widetilde 0|\cosh\frac{\widehat q}{2k}
\Biggl(1-ze^{-\frac{i}{2\hbar}\widehat q^2}
\frac{\widehat\Pi_+}{2\cosh\frac{\widehat p}{2}}
e^{\frac{i}{2\hbar}\widehat q^2}
\frac{\widehat\Pi_+}{2\cosh\frac{\widehat p}{2}}\Biggr)^{-1}|0\rangle,
\label{grandpot}
\end{align}
if we use the formula
\begin{align}
\det\Bigl(1-z\widehat{X}(1-|Y\rangle\langle\widetilde Y|)\Bigr)
=\det(1-z\widehat{X})\times
\langle\widetilde Y|(1-z\widehat{X})^{-1}|Y\rangle,
\end{align}
for a general operator $\widehat X$ and states $|Y\rangle$, $\langle\widetilde Y|$ normalized as $\langle\widetilde Y|Y\rangle=1$.

Before ending this subsection, we note that there is another way of rewriting the partition function \eqref{Z_k(N)-2} as the form \eqref{Z_k(N)-3}.
Namely, instead of \eqref{twoproj} we can define a different Hamiltonian by exchanging the projectors in \eqref{twoproj},
\begin{align}
e^{-\widehat H'}=e^{\frac{i}{2\hbar}\widehat q^2}
\biggl(\sinh\frac{\widehat q}{k}\biggr)^{-1}
\frac{\widehat\Pi_+}{2\cosh\frac{\widehat p}{2}}
\biggl(\sinh\frac{\widehat q}{k}\biggr)
e^{-\frac{i}{2\hbar}\widehat q^2}
\frac{\widehat\Pi_-}{2\cosh\frac{\widehat p}{2}},
\label{twoproj2}
\end{align}
to obtain the same expression \eqref{Z_k(N)-2} via \eqref{Z_k(N)-3}. 
However, one may notice that the expression (\ref{twoproj2}) is not as helpful as \eqref{twoproj} since the singularity of the matrix elements $\langle\mu|e^{-\widehat H'}|\nu\rangle$ (due to the presence of a hyperbolic sine function in the denominator) is not cancelled by the projectors.
In the following, we employ the Hamiltonian given by \eqref{twoproj}. 

\subsection{Even-parity determinant}
In this subsection we explain the computation of the determinant in \eqref{grandpot}.
After a similarity transformation,
\begin{align}
e^{-\frac{i}{2\hbar}\widehat p^2}
\Biggl[e^{-\frac{i}{2\hbar}\widehat q^2}
\frac{\widehat\Pi_+}{2\cosh\frac{\widehat p}{2}}
e^{\frac{i}{2\hbar}\widehat q^2}
\frac{\widehat\Pi_+}{2\cosh\frac{\widehat p}{2}}
\Biggr]e^{\frac{i}{2\hbar}\widehat p^2}
=\frac{1}{2\cosh\frac{\widehat q}{2}}
\frac{\widehat\Pi_+}{2\cosh\frac{\widehat p}{2}},
\end{align}
we find that the determinant is rewritten as
\begin{align}
&\det\Biggl(1-ze^{-\frac{i}{2\hbar}\widehat q^2}
\frac{\widehat\Pi_+}{2\cosh\frac{\widehat p}{2}}
e^{\frac{i}{2\hbar}\widehat q^2}
\frac{\widehat\Pi_+}{2\cosh\frac{\widehat p}{2}}\Biggr)
=\det\bigl(1-z\widehat\rho_+\bigr),
\label{detrho}
\end{align}
with the density matrix
\begin{align}
\widehat\rho_+=\frac{1}{\sqrt{2\cosh\frac{\widehat q}{2}}}
\frac{\widehat\Pi_+}{2\cosh\frac{\widehat p}{2}}
\frac{1}{\sqrt{2\cosh\frac{\widehat q}{2}}}.
\end{align}
The matrix element of the density matrix $\rho_+(\mu,\nu)=\langle\mu|\widehat\rho_+|\nu\rangle$ is simply the even-parity one
\begin{align}
\rho_+(\mu,\nu)
&=\frac{1}{k}
\frac{1}{\sqrt{2\cosh\frac{\mu}{2}}}
\frac{\cosh\frac{\mu}{2k}\cosh\frac{\nu}{2k}}
{\cosh\frac{\mu}{k}+\cosh\frac{\nu}{k}}
\frac{1}{\sqrt{2\cosh\frac{\nu}{2}}},
\end{align}
which has already been introduced in \cite{HMO1}.
The only difference is that, owing to the extra minus sign appearing in \eqref{scs}, the Fermi statistics is changed into the Bose statistics \eqref{detrho}.
See \cite{CM} for the appearance of the Bose statistics in the TBA equations.

As was already noted in \cite{HMO2}, similarly to the original density matrix of the ABJM theory, we can rewrite this density matrix in the form
\begin{align}
\rho_+(\mu,\nu)=\frac{E(\mu)E(\nu)}{M(\mu)+M(\nu)},
\end{align}
with
\begin{align}
E(\mu)=\frac{\cosh\frac{\mu}{2k}}{\sqrt{2\cosh\frac{\mu}{2}}},\quad
M(\mu)=k\cosh\frac{\mu}{k}.
\end{align}
This expression implies that we can compute the powers of the density matrix $\rho_+^n$ simply by selecting one vector $E$ and repeatedly multiplying it by the density matrix $\rho_+$,
\begin{align}
\rho_+^n(\mu,\nu)
=\sum_{m=0}^{n-1}(-1)^m
\frac{(\rho_+^mE)(\mu)(\rho_+^{n-1-m}E)(\nu)}
{M(\mu)-(-1)^nM(\nu)}.
\label{rhopower}
\end{align}
Then, the determinant \eqref{detrho} is obtained from $\rho_+^n(\mu,\nu)$ as 
\begin{align}
\det\left(1-z\widehat{\rho}_+\right)
=\exp\Biggl[-\sum_{n=1}^\infty\frac{z^n}{n}\tr\widehat{\rho}^n_+\Biggr],
\quad
\tr\widehat{\rho}^n_+=\int\frac{d\nu}{2\pi}\rho^n_+(\nu,\nu).
\end{align}

\subsection{Matrix elements}
Next, let us turn to the new term in \eqref{grandpot}.
As in the previous subsection, we perform the similarity transformation and rewrite each order of $z$ as
\begin{align}
&\langle\widetilde 0|\cosh\frac{\widehat q}{2k}
\Biggl[e^{-\frac{i}{2\hbar}\widehat q^2}
\frac{\widehat\Pi_+}{2\cosh\frac{\widehat p}{2}}
e^{\frac{i}{2\hbar}\widehat q^2}
\frac{1}{2\cosh\frac{\widehat p}{2}}\Biggr]^n|0\rangle
=\langle\widetilde 0|\cosh\frac{\widehat q}{2k}
e^{\frac{i}{2\hbar}\widehat p^2}
\Biggl[\frac{1}{2\cosh\frac{\widehat q}{2}}
\frac{\widehat\Pi_+}{2\cosh\frac{\widehat p}{2}}\Biggr]^n
e^{-\frac{i}{2\hbar}\widehat p^2}|0\rangle.
\end{align}
To study it, let us first compute the matrix elements
\begin{align}
\langle\widetilde 0|\cosh\frac{\widehat q}{2k}
e^{\frac{i}{2\hbar}\widehat p^2}|\mu\rangle
&=\int\frac{d\nu}{2\pi}\cosh\frac{\nu}{2k}
\frac{1}{\sqrt{-ik}}e^{-\frac{i}{2\hbar}(\nu-\mu)^2}
=e^{-\frac{\pi i}{4k}}\cosh\frac{\mu}{2k},
\nonumber\\
\langle\mu|\frac{1}{2\cosh\frac{\widehat q}{2}}
\frac{\widehat\Pi_+}{2\cosh\frac{\widehat p}{2}}|\nu\rangle
&=\frac{1}{k}\frac{1}{2\cosh\frac{\mu}{2}}
\frac{\cosh\frac{\mu}{2k}\cosh\frac{\nu}{2k}}
{\cosh\frac{\mu}{k}+\cosh\frac{\nu}{k}},
\nonumber\\
\langle\nu|e^{-\frac{i}{2\hbar}\widehat p^2}|0\rangle
&=\frac{1}{\sqrt{ik}}e^{\frac{i}{2\hbar}\nu^2},
\end{align}
where we have used
\begin{align}
\langle\nu|e^{\frac{i}{2\hbar}\widehat p^2}|\mu\rangle
=\frac{1}{\sqrt{-ik}}e^{-\frac{i}{2\hbar}(\nu-\mu)^2}.
\end{align}

Using these formulas, we find that the matrix elements are given by
\begin{align}
&\langle\widetilde 0|\cosh\frac{\widehat q}{2k}
e^{\frac{i}{2\hbar}\widehat p^2}
\Biggl[\frac{1}{2\cosh\frac{\widehat q}{2}}
\frac{\widehat\Pi_+}{2\cosh\frac{\widehat p}{2}}\Biggr]^n
e^{-\frac{i}{2\hbar}\widehat p^2}|0\rangle
=\int\frac{d\nu}{2\pi}\frac{(\rho_+^nE)(\nu)}{E(\nu)}F(\nu),
\label{elem}
\end{align}
with
\begin{align}
F(\nu)=\frac{e^{-\frac{\pi i}{4k}}}{\sqrt{ik}}
\cosh\frac{\nu}{2k}e^{\frac{i}{2\hbar}\nu^2}.
\end{align}
Note that in the computation of either the determinant using \eqref{rhopower} in the previous subsection or the matrix elements using \eqref{elem} in this subsection, our starting point is the same.
We can select one vector $E$ and repeatedly multiply it by the density matrix $\rho_+$.

In the last step of the computation, we must perform integrations of the form
\begin{align}
\int d\nu
\frac{e^{\frac{n_1+n_2}{2}\nu}
(\cosh{\frac{m_1}{2k}\nu})(\sinh{\frac{m_2}{2k}\nu})\nu^p}
{(e^\nu+1)^{n_1}(e^\nu-1)^{n_2}}e^{\frac i{4\pi k}\nu^2},
\end{align}
for non-negative integers $n_1$, $n_2$, $m_1$, $m_2$ and $p$.
For the integration, we first apply the partial fraction decomposition to make either $n_1$ or $n_2$ equal to zero.
For the case of non-vanishing $n_2$, we can shift the integration variable $\nu$ by $\pi i$ so that the integration reduces to the case of non-vanishing $n_1$.
Finally, all we have to consider is the integration of
\begin{align}
I^{m,p}_n=\int d\nu
\frac{e^{\frac m{2k}\nu}\nu^p}{(e^\nu+1)^n}e^{\frac i{4\pi k}\nu^2}.
\label{integral}
\end{align}
The integration formulas for $I^{m,p}_n$ are given in appendix \ref{inteven}.

\subsection{Numerical results}
Using the Fermi gas formalism we have established, we find the first few exact values of the partition function $Z_k(N)$ via \eqref{GCPF}.
We first observe that the complex phase is given by
\begin{align}
\frac{Z_k(N)}{|Z_k(N)|}=(-i)^N,
\label{phase}
\end{align}
for integral $k$.
We have computed up to $N_\text{max}=7,8,4,7,3,7,3,3$ for $k=1,2,3,4,5,6,8,12$, respectively, and we list the first few exact values in table \ref{values}.\footnote{Formulas for $N=1,2$ and a general value $k$ have been given by Tomoki Nosaka.
We have checked their consistency with table \ref{values}.
We are grateful to him for sharing his results with us.}

\begin{table}[!p]
\begin{align*}
&|Z_1(1)|=\frac{1}{4\sqrt{2}},\quad
|Z_1(2)|=\frac{-2+\pi}{64\pi},\quad
|Z_1(3)|=\frac{-2\sqrt{2}+(8-5\sqrt{2})\pi}{512\pi},\\
&\quad
|Z_1(4)|=\frac{36-4\pi+(-25+16\sqrt{2})\pi^2}{8192\pi^2},\\
&|Z_2(1)|=\frac{1}{16},\quad
|Z_2(2)|=\frac{-8+\pi^2}{512\pi^2},\quad
|Z_2(3)|=\frac{-8-32\pi+11\pi^2}{8192\pi^2},\\
&\quad
|Z_2(4)|=\frac{192-560\pi^2-384\pi^3+177\pi^4}{1572864\pi^4},\\
&|Z_3(1)|=\frac{3\sqrt{2}-2\sqrt{3}}{24},\quad
|Z_3(2)|=\frac{6+(3-2\sqrt{6})\pi}{192\pi},\\
&\quad
|Z_3(3)|=\frac{54\sqrt{2}+60\sqrt{3}+(40-135\sqrt{2}+54\sqrt{3})\pi}
{13824\pi},\\
&\quad
|Z_3(4)|
=\frac{1188+36(9+10\sqrt{6})\pi
+(-4721+240\sqrt{2}+864\sqrt{3}+972\sqrt{6})\pi^2}{663552\pi^2},\\
&|Z_4(1)|=\frac{1}{16\pi},\quad
|Z_4(2)|=\frac{10-\pi^2}{1024\pi^2},\quad
|Z_4(3)|=\frac{78-121\pi^2+36\pi^3}{147456\pi^3},\\
&\quad
|Z_4(4)|=\frac{876-4148\pi^2-2016\pi^3+1053\pi^4}{18874368\pi^4},\\
&|Z_5(1)|=\frac{-5+2\sqrt{5+\sqrt{5}}}{20\sqrt{2}},\quad
|Z_5(2)|=\frac{-10+(5+4\sqrt{5}-4\sqrt{5+\sqrt{5}})\pi}{320\pi},\\
&\quad
|Z_5(3)|=\frac{250+60\sqrt{5+\sqrt{5}}+(625+52\sqrt{2}-100\sqrt{5}-64\sqrt{10}-150\sqrt{5+\sqrt{5}})\pi}
{32000\sqrt{2}\pi},\\
&|Z_6(1)|=\frac{-3+2\sqrt{3}}{48},\quad
|Z_6(2)|=\frac{-216+(209-108\sqrt{3})\pi^2}{41472\pi^2},\\
&\quad
|Z_6(3)|=\frac{(1944-4752\sqrt{3})+1440\pi+(-9171+5398\sqrt{3})\pi^2}
{5971968\pi^2},\\
&|Z_8(1)|=\frac{-2+\pi}{64\pi},\quad
|Z_8(2)|=\frac{36-4\pi+(-25+16\sqrt{2})\pi^2}{8192\pi^2},\\
&\quad
|Z_8(3)|=\frac{-600+324\pi+(2242+2016\sqrt{2})\pi^2+(-1431-144\sqrt{2})\pi^3}{4718592\pi^3},\\
&|Z_{12}(1)|=\frac{9-\sqrt{3}\pi}{432\pi},\quad
|Z_{12}(2)|=\frac{702-12\sqrt{3}\pi+(-1561+864\sqrt{3})\pi^2}
{248832\pi^2},\\
&\quad
|Z_{12}(3)|=\frac{17982-2106\sqrt{3}\pi-27(4585+864\sqrt{3})\pi^2
+(45684+3679\sqrt{3})\pi^3}{322486272\pi^3}.
\end{align*}
\caption{Exact values of the partition function $|Z_k(N)|$ of the OSp theory.}
\label{values}
\end{table}

After finding these exact values of the partition function, we can read off the perturbative and non-perturbative coefficients of the grand potential $J_k(\mu)$ \eqref{totalgrandpot} by numerical fitting.
As in the analysis \cite{MM} of the ABJ theory \cite{HLLLP2,ABJ}, the partition function has a simple phase dependence \eqref{phase} and we define the grand potential $J_k(\mu)$ by taking absolute values.
Details of the numerical fitting can be found in \cite{PTEP}, which we only explain briefly in the following.
We make an ansatz for the grand potential $J_k(\mu)=J^\text{(pert)}_k(\mu)+J^\text{(np)}_k(\mu)$,
\begin{align}
J^\text{(pert)}_k(\mu)=\frac{C_k}3\mu^3+B_k\mu+A_k,\quad
J^\text{(np)}_k(\mu)=\sum_{a}f_a(\mu)e^{-a\mu},
\label{ansatz-np}
\end{align}
with polynomials $f_a(\mu)$, and find the best fit for the coefficients of the polynomials $f_a(\mu)$ by comparing the absolute values of the partition function $|Z_k(N)|$ in table \ref{values} with the inverse transformation of \eqref{totalgrandpot},
\begin{align}
|Z_k(N)|=\int_{-\infty i}^{+\infty i}
\frac{d\mu}{2\pi i}e^{J_k(\mu)-\mu N}.
\label{fitting}
\end{align}
Then, the perturbative coefficients are found to be given by 
\begin{align}
C^\text{OSp}_k=\frac{1}{\pi^2k},\quad
B^\text{OSp}_k=\frac{5}{12k}+\frac{k}{48},\quad
A^\text{OSp}_k=\frac{1}{2}(A^\text{ABJM}_k-\log 2),
\end{align}
while the non-perturbative coefficients are summarized in table \ref{instanton}.
It is interesting to observe that when $k=1$, $J^\text{OSp}_1(\mu)$ coincides with the grand potential constructed from $\det(1+z\rho_+)$ in \cite{HMO1,GHM2}.

\begin{table}[!p]
\begin{align*}
&J^\text{(np)}_{1}
=\frac{1}{\sqrt{2}}e^{-\mu}
-\frac{4}{3\sqrt{2}}e^{-3\mu}
+\biggl[\frac{2\mu^2+\mu/2+1/8}{\pi^2}\biggr]e^{-4\mu}
-\frac{16}{5\sqrt{2}}e^{-5\mu}+\frac{64}{7\sqrt{2}}e^{-7\mu}
\\&\quad
+\biggl[-\frac{13\mu^2+\mu/8+9/64}{\pi^2}+2\biggr]e^{-8\mu}
+\frac{256}{9\sqrt{2}}e^{-9\mu}-\frac{1024}{11\sqrt{2}}e^{-11\mu}
\\&\quad
+\biggl[\frac{368\mu^2-76\mu/3+77/36}{3\pi^2}-32\biggr]e^{-12\mu}
-\frac{4096}{13\sqrt{2}}e^{-13\mu}+\frac{16384}{15\sqrt{2}}e^{-15\mu}
+{\cal O}(e^{-16\mu}),\\
&J^\text{(np)}_{2}=e^{-\mu}
+\biggl[-\frac{2\mu^2+\mu+1/2}{\pi^2}\biggr]e^{-2\mu}
+\frac{16}{3}e^{-3\mu}
+\biggl[-\frac{13\mu^2+\mu/4+9/16}{\pi^2}+2\biggr]e^{-4\mu}
\\&\quad
+\frac{256}{5}e^{-5\mu}
+\biggl[-\frac{368\mu^2-152\mu/3+77/9}{3\pi^2}+32\biggr]e^{-6\mu}
+\frac{4096}{7}e^{-7\mu}+{\cal O}(e^{-8\mu}),\\
&J^\text{(np)}_{3}=\frac{1}{\sqrt{2}}e^{-\mu}
-\frac{1}{3}e^{-\frac{4}{3}\mu}
+\frac{1}{2}e^{-\frac{8}{3}\mu}
-\frac{4}{3\sqrt{2}}e^{-3\mu}
+\biggl[\frac{2\mu^2+\mu/2+1/8}{3\pi^2}-\frac{8}{9}\biggr]e^{-4\mu}
-\frac{16}{5\sqrt{2}}e^{-5\mu}
\\&\quad
+\frac{25}{36}e^{-\frac{16}{3}\mu}
-\frac{12}{5}e^{-\frac{20}{3}\mu}
+\frac{64}{7\sqrt{2}}e^{-7\mu}
+\biggl[-\frac{13\mu^2+\mu/8+9/64}{3\pi^2}+\frac{88}{9}\biggr]e^{-8\mu}
+\frac{256}{9\sqrt{2}}e^{-9\mu}
\\&\quad
-\frac{947}{189}e^{-\frac{28}{3}\mu}
+{\cal O}(e^{-\frac{32}{3}\mu}),\\
&J^\text{(np)}_{4}=\biggl[\frac{2\mu^2+\mu+1/2}{2\pi^2}\biggr]e^{-2\mu}
+\biggl[-\frac{13\mu^2+\mu/4+9/16}{2\pi^2}+2\biggr]e^{-4\mu}
\\&\quad
+\biggl[\frac{184\mu^2-76\mu/3+77/18}{3\pi^2}-32\biggr]e^{-6\mu}
+{\cal O}(e^{-8\mu}),\\
&J^\text{(np)}_{5}=\frac{-5+3\sqrt{5}}{10}e^{-\frac{4}{5}\mu}
-\frac{1}{\sqrt{2}}e^{-\mu}
+\frac{5+7\sqrt{5}}{20}e^{-\frac{8}{5}\mu}
+\frac{5-3\sqrt{5}}{15}e^{-\frac{12}{5}\mu}
+\frac{4}{3\sqrt{2}}e^{-3\mu}
\\&\quad
+\frac{5-31\sqrt{5}}{40}e^{-\frac{16}{5}\mu}
+\biggl[\frac{2\mu^2+\mu/2+1/8}{5\pi^2}
-\frac{1+5\sqrt{5}}{10}\biggr]e^{-4\mu}
+\frac{485-91\sqrt{5}}{150}e^{-\frac{24}{5}\mu}
\\&\quad
+\frac{16}{5\sqrt{2}}e^{-5\mu}
+\frac{170+263\sqrt{5}}{175}e^{-\frac{28}{5}\mu}
+{\cal O}(e^{-\frac{32}{5}\mu}),\\
&J^\text{(np)}_{6}=\frac{1}{3}e^{-\frac{2}{3}\mu}
-e^{-\mu}+\frac{1}{2}e^{-\frac{4}{3}\mu}
+\biggl[-\frac{2\mu^2+\mu+1/2}{3\pi^2}+\frac{8}{9}\biggr]e^{-2\mu}
+\frac{25}{36}e^{-\frac{8}{3}\mu}
-\frac{16}{3}e^{-3\mu}
\\&\quad
+\frac{12}{5}e^{-\frac{10}{3}\mu}
+\biggl[-\frac{13\mu^2+\mu/4+9/16}{3\pi^2}+\frac{88}{9}\biggr]e^{-4\mu}
+\frac{947}{189}e^{-\frac{14}{3}\mu}-\frac{256}{5}e^{-5\mu}
+{\cal O}(e^{-\frac{16}{3}\mu}),\\
&J^\text{(np)}_{8}=\frac{\sqrt{2}}{2}e^{-\frac{1}{2}\mu}
-\frac{2\sqrt{2}}{3}e^{-\frac{3}{2}\mu}
+\biggl[\frac{2\mu^2+\mu+1/2}{4\pi^2}\biggr]e^{-2\mu}
-\frac{8\sqrt{2}}{5}e^{-\frac{5}{2}\mu}
+\frac{32\sqrt{2}}{7}e^{-\frac{7}{2}\mu}
+{\cal O}(e^{-4\mu}),\\
&J^\text{(np)}_{12}=\sqrt{3}e^{-\frac{1}{3}\mu}
-\frac{7}{6}e^{-\frac{2}{3}\mu}
+\frac{9}{4}e^{-\frac{4}{3}\mu}
-\frac{16\sqrt{3}}{5}e^{-\frac{5}{3}\mu}
+\biggl[\frac{2\mu^2+\mu+1/2}{6\pi^2}+\frac{74}{9}\biggr]e^{-2\mu}
\\&\quad
-\frac{185\sqrt{3}}{21}e^{-\frac{7}{3}\mu}
+{\cal O}(e^{-\frac{8}{3}\mu}).
\end{align*}
\caption{Non-perturbative terms of the grand potential $J^\text{(np)}_k(\mu)$ of the OSp theory.}
\label{instanton}
\end{table}

If we compare the results in table \ref{instanton} with those of the ABJM theory, we can make the following observations.
\begin{itemize}
\item
For the perturbative part, the coefficients in the two theories are related by
\begin{align}
C^\text{OSp}_k=\frac{C^\text{ABJM}_k}{2},\quad
B^\text{OSp}_k=\frac{B^\text{ABJM}_k+\pi^2C^\text{ABJM}_k/4}{2},\quad
A^\text{OSp}_k=\frac{A^\text{ABJM}_k-\log 2}{2}.
\end{align}
\item
The non-perturbative part consists of exponents $e^{-\frac{4m}{k}\mu}$ and $e^{-\ell\mu}$ ($m,\ell\in{\mathbb N}$).
These two exponents are identified as the worldsheet instanton and the membrane instanton, respectively, from their $k$-dependences.
\item
Compared with the membrane instanton of the ABJM theory, where the exponents are $e^{-2\ell\mu}$, we also encounter odd powers of $e^{-\mu}$ for the OSp theory.
However, the coefficients of these terms are quite simple,\footnote{Owing to the coincidence noted above, $J^\text{OSp (odd MB)}_{k=1}(\mu)$ in \eqref{mod4} has already appeared in \cite{GHM2}.
Although the results for $k\equiv 7$ mod $8$ are missing and we only have one result for $k\equiv 1,2,3,5,6$ mod $8$, we believe that our formula \eqref{mod4} is a natural extrapolation.}
\begin{align}
&J^\text{OSp (odd MB)}_{k\equiv 1,3\,\text{mod}\,8}(\mu)
=-J^\text{OSp (odd MB)}_{k\equiv 5,7\,\text{mod}\,8}(\mu)
=\frac{1}{8}\log\frac{1+2\sqrt{2}e^{-\mu}+4e^{-2\mu}}
{1-2\sqrt{2}e^{-\mu}+4e^{-2\mu}},\nonumber\\
&J^\text{OSp (odd MB)}_{k\equiv 2\,\text{mod}\,8}(\mu)
=-J^\text{OSp (odd MB)}_{k\equiv 6\,\text{mod}\,8}(\mu)
=\frac{1}{8}\log\frac{1+4e^{-\mu}}{1-4e^{-\mu}},\nonumber\\
&J^\text{OSp (odd MB)}_{k\equiv 0\,\text{mod}\,4}(\mu)=0.
\label{mod4}
\end{align}
\item
The coefficient polynomials of $e^{-2\ell\mu}$ proportional to $\pi^{-2}$ are the same for the OSp theory and the ABJM theory except for an extra relative factor $2(-1)^\ell$.
\end{itemize}

\subsection{From OSp to ABJM}\label{fromto}
In the previous subsection, we found an explicit expression for the grand potential for $k=1,2,3,4,5,6,8,12$ and made several observations on the exponents and the coefficients.

These observations suggest that the argument $\mu\pm\pi i/2$ should be considered.\footnote{We are grateful to Tomoki Nosaka for valuable discussions on this point.}
In fact, it is not difficult to find that
\begin{align}
J_k^\text{OSp(pert)}\Bigl(\mu+\frac{\pi i}{2}\Bigr)
+J_k^\text{OSp(pert)}\Bigl(\mu-\frac{\pi i}{2}\Bigr)
=J_k^\text{ABJM(pert)}(\mu)-\log 2,
\label{pert}
\end{align}
and notice that, when generalizing the above equation non-perturbatively, the extra non-perturbative $e^{-\ell\mu}$ terms with odd $\ell$ on the left-hand side cancel among themselves.

As reviewed in the introduction, the analysis in \cite{HM} implies that the grand potential has an intricate constraint and, in particular, it is not appropriate to consider only the sum of two terms such as in the above equation.
Instead, we can study the duplicate quiver using \eqref{2repeat}.
Surprisingly, we find that the result correctly reproduces the grand potential of the ABJM theory as in \eqref{OSpABJM},
\begin{align}
J^\text{OSp[2]}_k(\mu)=J^\text{ABJM[1]}_k(\mu/2)-\log 2,
\label{full}
\end{align}
including the non-perturbative terms.
For a comparison, see appendix A in \cite{HMO3}.

Since the partition function of the ABJM theory is known, from this observation, we can solve the whole non-perturbative expansion except for the $e^{-\ell\mu}$ terms with odd $\ell$.
Combining this with the observation \eqref{mod4}, we claim that we have solved the partition function of the OSp theory for all integers $k$.

Note that, since in \cite{HM} the formula \eqref{2repeat} is derived under the assumption $Z_k(N)>0$, it must be applied with care to the OSp theory because of the phase factor \eqref{phase}. 
Because of this phase factor and the rescaling by a factor of $2$, the relation \eqref{full} does not literally mean that the grand potential of the duplicate OSp theory is the same as that of the ABJM theory.
Instead, the grand partition function of the duplicate OSp theory is 
\begin{align}
\Biggl[\sum_{N=0}^\infty Z^\text{OSp}_k(N)e^{\frac{\mu+\pi i}{2}N}\Biggr]
\Biggl[\sum_{N=0}^\infty Z^\text{OSp}_k(N)e^{\frac{\mu-\pi i}{2}N}\Biggr]
=\sum_{N=0}^\infty(-1)^NZ^\text{ABJM}_k(2N)e^{\mu N},
\label{gcpf}
\end{align}
which is not that of the ABJM theory. 
See figures 9--13 in \cite{HMO2} for comparison.
Then, it is natural to ask the meaning of $J^\text{OSp[2]}_k(\mu)$.
Using \eqref{full}, we can show
\begin{equation}
\sum_{n=-\infty}^\infty e^{J^\text{OSp[2]}_k(\mu+2\pi in)}
=\sum_{N=0}^\infty Z_k^{\rm ABJM}(2N)e^{\mu N}, 
\end{equation}
indicating that $J^\text{OSp[2]}_k(\mu)$ is indeed the grand potential of the duplicate OSp theory if we define it by dropping the phase factor in \eqref{gcpf} as before.

\section{Summary and discussion}
We have studied the partition function of the OSp theory, which is the ${\cal N}=5$ supersymmetric orthosymplectic Chern-Simons theory with the gauge group O$(2N)\times$USp(2N).
Its non-perturbative structure was clarified by considering the duplication of the quiver.

We list a few technical issues that remain unresolved in solving the OSp theory, and then discuss some future directions.

First, the WKB expansion of the OSp theory is missing.
We directly proceeded to the evaluation of the partition function but it is desirable to also study the WKB expansion to obtain a full understanding of the membrane instantons.
For example, we do not know how the odd membrane instantons behave for non-integral $k$.

Secondly, we have followed the standard method to obtain the non-perturbative expansion of the grand potential and substituted the results into the repetition formula \eqref{2repeat} to find the relation \eqref{OSpABJM} between the OSp theory and the ABJM theory.
However, it would be interesting to attempt to prove the relation directly from the density matrix in \eqref{twoproj}.
A direct proof would lead to more general relations between the families of orientifold theories and those of original theories.

Here we have only studied the partition function of the OSp theory with equal ranks.
As in the ABJM theory, it would be interesting to study the BPS Wilson loop in the theory and/or the ABJ deformation O$(2N_1)\times$USp$(2N_2)$.
A particularly interesting question is when the BPS Wilson loop reduces to the character of the OSp$(2N_1|2N_2)$ supergroup in the localization.
For the OSp character, we hope that the formalism in \cite{HHMO,MM} will directly apply to the OSp theory.
Using this formalism, among others, we hope to establish whether our relation between the OSp theory and the ABJM theory \eqref{OSpABJM} holds for the ABJ deformation.
Also, it will be interesting to study the OSp theory in terms of topological string theory as in the ABJM theory.
For example, study from the viewpoint of the dual spectral problem \cite{KM,GHM1,WZH,CGM} may also be worthwhile.
We would like to pursue these directions following our previous attempts \cite{JP,TM}.

Technically, the Fermi gas Hamiltonians in the OSp matrix model and the $\widehat D$-type quiver matrix model \cite{ADF,MN4} are similar.
It would be interesting to determine the relation between the two Hamiltonians.

Let us now return to our original question of the constraint on the partition function.
Partition functions in physical theories often have a certain kind of modular invariance.
The most standard example is, of course, the one-loop string worldsheet partition function, which is expressed by modular invariant combinations of theta functions.
Owing to the constraints imposed by the modular invariance, we cannot construct a new theory simply by projecting out some excitation modes.
In addition to the untwisted sector, which is obtained simply from the orbifold or orientifold projection, we need to add twisted sectors to recover the broken modular invariance.

Similarly, the expression for the supersymmetric Chern-Simons theory is considered to be strongly constrained.
In parallel to the worldsheet partition function we have called the $n=0$ part in \eqref{2repeat} the untwisted sector and the $n\ne 0$ part the twisted sector.
This similarity led us to \eqref{full} from \eqref{pert}.
However, we do not have a clear picture of the modular invariance behind the constraint.
A study of the modular invariance in five-dimensional partition functions \cite{LV} or in multiple sine functions \cite{Narukawa} may increase our understanding of the matrix model studied in this work.

\section*{Acknowledgements}
We are grateful to Tomoki Nosaka for collaborative discussions at various stages of this work.
We would also like to thank Benjamin Assel, Nadav Drukker, Yasuyuki Hatsuda, Kimyeong Lee, Takuya Matsumoto, Takahiro Nishinaka, Kazumi Okuyama and Jaemo Park for valuable discussions.
The work of S.M.\ was supported by JSPS Grant-in-Aid for Scientific
Research (C) \# 26400245.
S.M.\ would like to thank Yukawa Institute for Theoretical Physics at Kyoto University for hospitality.

\appendix

\section{Integrals} \label{inteven}
To compute the grand potential at each order of $z$ exactly, we need to evaluate the integrals $I^{m,p}_n$ given by \eqref{integral}.
Using integration by parts, we easily obtain the recursive formula
\begin{align}
I^{m,p}_n=2\pi ik(p-1)I^{m,p-2}_n
-\pi i(2kn-m)I^{m,p-1}_n+2\pi iknI^{m,p-1}_{n+1},
\end{align}
which reduces the evaluation of $I^{m,p}_n$ to those with lower $p$.
Hence, in the following, we can concentrate on the evaluation of $I^{m,0}_n$.

\subsection{Odd $k$}
For the evaluation of $I^{m,0}_n$ with even $m$, we first rewrite the integrand using
\begin{align}
\frac{1}{(e^\nu+1)^n}=\frac{e^\nu+1}{(e^\nu+1)^{n+1}}.
\end{align}
Then, we find that the $e^\nu$ term is simply the same integral with the integration contour inversed and shifted by $2\pi ik$,
\begin{align}
\int_\mathbb{R}d\nu
\frac{e^\nu e^{\frac m{2k}\nu}}{(e^\nu+1)^{n+1}}e^{\frac i{4\pi k}\nu^2}
=-\int_{\mathbb{R}-2\pi ik}d\nu
\frac{e^{\frac m{2k}\nu}}{(e^\nu+1)^{n+1}}e^{\frac i{4\pi k}\nu^2}.
\end{align}
Hence, we can evaluate the integration by extracting the residues between ${\mathbb R}$ and ${\mathbb R}-2\pi ik$,
\begin{align}
I^{m,0}_n=-2\pi i\sum_{l=1}^k\Res_{-(2l-1)\pi i}
\frac{e^{\frac m{2k}\nu}}{(e^\nu+1)^{n+1}}e^{\frac i{4\pi k}\nu^2}.
\end{align}

If $m$ is odd, on the other hand, the same technique of reversing and shifting by $2\pi ik$ leads to
\begin{align}
\int_\mathbb{R}d\nu
\frac{(-e^\nu+1)e^{\frac m{2k}\nu}}{(e^\nu+1)^{n+1}}
e^{\frac i{4\pi k}\nu^2}
=-2\pi i\sum_{l=1}^k\Res_{-(2l-1)\pi i}
\frac{e^{\frac m{2k}\nu}}{(e^\nu+1)^{n+1}}e^{\frac i{4\pi k}\nu^2},
\end{align}
which implies the following recursive formula for $n$:
\begin{align}
I^{m,0}_n=\frac12I^{m,0}_{n-1}
-\pi i\sum_{l=1}^k\Res_{-(2l-1)\pi i}
\frac{e^{\frac m{2k}\nu}}{(e^\nu+1)^{n}}e^{\frac i{4\pi k}\nu^2}. 
\end{align}
This reduces the evaluation of $I^{m,0}_n$ to that of $I^{m,0}_0$, which can be easily performed to give
\begin{align}
I^{m,0}_0=2\pi\sqrt{ik}\,e^{\frac{\pi i}{4k}m^2}.
\end{align} 

\subsection{Even $k$}
When $k$ is even, it turns out that we only need to evaluate integrals of the form 
\begin{align}
\int_\mathbb{R}d\nu\frac{e^{\frac{2m+1}{2k}\nu}}{(\cosh\frac\nu2)^n}e^{\frac i{4\pi k}\nu^2}=2^nI^{2m+nk+1,0}_n. 
\end{align}
Therefore, it is sufficient to evaluate $I^{m,0}_n$ with odd $m$. 
We apply the same technique and obtain
\begin{align}
I^{m,0}_n=-2\pi i\sum_{l=1}^k\Res_{-(2l-1)\pi i}
\frac{e^{\frac m{2k}\nu}}{(e^\nu+1)^{n+1}}e^{\frac i{4\pi k}\nu^2}.
\end{align}

\end{document}